\documentclass[12pt, epsfig]{article}
\usepackage{amsmath,amsfonts,amssymb,amsthm,amstext,amscd,eucal}
\usepackage[all]{xy}
\usepackage{dsfont}
\usepackage{hyperref}
\usepackage{amsmath}
\usepackage{slashed}
\usepackage[dvips]{color}
\usepackage{color}
\usepackage[active]{srcltx}
\usepackage{enumitem}
\setlist{nolistsep}

\makeatletter \@addtoreset{equation}{section}

\makeatletter\renewcommand\section{\@startsection {section}{1}{\z@}%
                                   {-3.5ex \@plus -1ex \@minus -.2ex}
                                   {2.3ex \@plus.2ex}%
                                   {\normalfont\large\bfseries}}
\renewcommand\subsection{\@startsection{subsection}{2}{\z@}%
                                     {-3.25ex\@plus -1ex \@minus -.2ex}%
                                     {1.5ex \@plus .2ex}%
                                     {\normalfont\bfseries}}

\newcommand{\be}{\begin{equation}}
\newcommand{\ee}{\end{equation}}
\newcommand{\bea}{\begin{eqnarray}}
\newcommand{\eea}{\end{eqnarray}}
\newcommand{\bse}{\begin{subequations}}
\newcommand{\ese}{\end{subequations}}
\newcommand{\beqa}{\begin{eqnarray}}
\newcommand{\eeqa}{\end{eqnarray}}
\newcommand{\beqar}{\begin{eqnarray*}}
\newcommand{\eeqar}{\end{eqnarray*}}
\newcommand{\bi}{\begin{itemize}}
\newcommand{\ei}{\end{itemize}}
\newcommand{\bn}{\begin{enumerate}}
\newcommand{\en}{\end{enumerate}}
\newcommand{\fixme}[1]{\textbf{FIXME: }$\langle$\textit{#1}$\rangle$}
\newcommand{\note}[1]{\textbf{NOTE: }$\langle$\textit{#1}$\rangle$}
\newcommand{\ba}{\begin{array}}
\newcommand{\ea}{\end{array}}
\newcommand{\bc}{\begin{center}}
\newcommand{\ec}{\end{center}}

\definecolor{darkgreen}{rgb}{0,0.3,0}
\definecolor{darkblue}{rgb}{0,0,0.3}
\definecolor{darkred}{rgb}{0.7,0,0}

\begin{document}

\begin{titlepage}

\begin{flushright}\vspace{-3cm}
{
IPM/P-2014/013  \\
\today }\end{flushright}
\vspace{-.5cm}

\begin{center}
\Large{\bf{On Quantization of AdS$_3$ Gravity I: \\
\textit{Semi-Classical Analysis}}} \vspace{4mm}

\large{\bf{M.M. Sheikh-Jabbari\footnote{e-mail:
jabbari@theory.ipm.ac.ir}$^{;\ a,b}$
 and  H. Yavartanoo\footnote{e-mail: yavar@itp.ac.cn }$^{;\ c}$  }}
\\

\vspace{5mm}
\normalsize
\bigskip\medskip
{$^a$ \it School of Physics, Institute for Research in Fundamental
Sciences (IPM),\\ P.O.Box 19395-5531, Tehran, Iran}\\
{$^b$ \it  Department of Physics, Kyung Hee University, Seoul 130-701, Korea}\\
\smallskip
{$^c$ \it  State Key Laboratory of Theoretical Physics, Institute of Theoretical Physics,
Chinese Academy of Sciences, Beijing 100190, China.
 }\\

\end{center}
\setcounter{footnote}{0}

\begin{abstract}
\noindent
In this work we explore ideas in quantizing AdS$_3$ Einstein gravity. We start with the most general solution to the 3d gravity theory which respects Brown-Henneaux boundary
conditions. These solutions are specified by two holomorphic functions and satisfy simple superposition rule.
These geometries  generically have a bifurcate Killing horizon (with a noncompact or not simply connected bifurcation curve) which is not an event horizon. Nonetheless, there are superpositions of these geometries which have event horizon.
We propose to view these geometries as ``semiclassical fuzzball microstates'' of  BTZ black holes appearing as superposition of these geometries.
The details of quantization of these semiclassical microstates will be discussed in an upcoming work.

\end{abstract}

\end{titlepage}
\renewcommand{\baselinestretch}{1.1}  
\tableofcontents

\section{Introduction}

Three dimensional gravity  has been long used as a laboratory for testing quantum gravity ideas \cite{3d-gravity,{Deser-Jackiw},{AdS3-gravity-Witten}}.
In three dimensions Einstein gravity is classically equivalent to a Chern-Simons gauge theory and has no propagating degrees of freedom. Nonetheless,
it still has non-trivial (quantum and topological) features. In presence of a negative cosmological constant, the AdS$_3$ gravity, the theory seems to show more interesting features than in the asymptotic flat space, e.g. it admits black hole solutions \cite{BTZ,BTZ-SDA}.\footnote{There are however, interesting recent works on 3d gravity in flat space, and the BMS$_3$ as its asymptotic symmetry group. See e.g. \cite{BT,BGG-solution} and references therein.} In this work we will be revisiting the AdS$_3$ gravity and study some features of
its generic solutions.

It is well known that all solutions to AdS$_3$  gravity, which satisfy
\be\label{3d-Einstein}
R_{\mu\nu}=-\frac{2}{\ell^2} g_{\mu\nu}\,,
\ee
are locally AdS$_3$ and are hence  locally invariant under $sl(2,R)\times sl(2,R)$.
The ``global AdS$_3$ geometry'' is the only one which has this isometry globally, i.e. it is invariant under global $Sl(2,R)\times Sl(2,R)$ isometry group. Any given solution to \eqref{3d-Einstein} can then be classified by its
\emph{global} isometry group. A class of these geometries are obtained through orbifolding of AdS$_3$ by an appropriate subgroup of $Sl(2,R)\times Sl(2,R)$
and fall into three categories: conic spaces \cite{Deser-Jackiw}, BTZ black holes \cite{BTZ}, extremal BTZ \cite{BTZ} and self-dual orbifold \cite{BTZ, BTZ-SDA}.
These geometries are all asymptotically AdS$_3$, but have different global isometry group. Conic spaces have $Sl(2,R)/Z_k\times Sl(2,R)/Z_l$ with $k,l\in \mathbb{Z}$
global isometry group, BTZ black hole have $U(1)_+\times U(1)_-$ global isometry, with compact $U(1)$'s. The self-dual orbifold has global $Sl(2,R)\times U(1)$, where the $U(1)$
is compact and the Abelian subgroup of $Sl(2,R)$ is noncompact.

The most general class of solution to \eqref{3d-Einstein} which satisfy Brown-Henneaux boundary conditions \cite{Brown-Henneaux} (Brown-Henneaux near boundary fall off behavior) can be written in the Fefferman-Graham coordinates
\cite{Fefferman:2007rka} as \cite{Banados,{Sol-Sken}}
\be\label{generic-solutions}
ds^2=\frac{\ell^2 dr^2}{r^2}-(rdx^+-\frac{\ell^2}{r}f_-dx^-)(rdx^--\frac{\ell^2}{r}f_+dx^+)\,,
\ee
where $f_{+}=f_+(x^+)$  and $f_{-}=f_-(x^-)$ are arbitrary periodic functions.
Depending on whether $x^+$ or $x^-$ are compact or not, one can distinguish three different cases:
$\vspace{-0.1cm}$
\bn
\item  both $x^\pm$ are compact with period $2\pi$, more explicitly, $x^\pm=t/\ell\pm\varphi$ and $\varphi\in[0,2\pi]$;
\item $x^+$ is noncompact $x^+\in \mathbb{R}$ while $x^-$ is compact, $x^-\in [0,2\pi]$;
\item neither of $x^\pm$ are compact.
$\vspace{-0.1cm}$
\en
As we will show in the last case (with noncompact $x^\pm$) there is no obstruction in removing functions $f_\pm$ and  mapping the geometry to
AdS$_3$ in Poincar\'{e} patch using the local coordinate transformations. We will therefore only focus on the first two cases with at least one compact direction.
We will discuss that these two cases correspond to distinct physical cases and will hence be discussed separately.

The case with both $x^\pm$ compact, as we will show,  is correlated with the BTZ black hole ``semiclassical microstates''. This case will be discussed in sections \ref{sec-2} and \ref{sec-3}. When $f_+$ (or $f_-$) is zero, one can consider $x^+$ (or $x^-$) to be compact or not. The case with only $x^-$ compact corresponds to geometries whose near horizon limit leads to self-dual AdS$_3$ orbifold. The class of solutions with $f_+=0$, which will be dealt with  in section \ref{sec-4}, are those constructed and analyzed in \cite{LL-paper}.

The geometries in \eqref{generic-solutions} are diffeomorphic to AdS$_3$ and their large $r$ deviation  from the AdS$_3$ geometry, when both $x^\pm$ are compact, is given by Brown-Henneaux diffeomorphisms.
These diffeomorphisms are specified by the same holomorphic or anti-holomorphic $f_\pm$ functions. In other words, and as we will show explicitly in the next section,
\eqref{generic-solutions} are in
one-to-one correspondence with Brown-Henneaux diffeomorphisms around BTZ black holes or conic spaces and extend the Brown-Henneaux boundary gravitons to the whole bulk.
As discuss in \cite{BGG-solution}, one can associate Virasoro charges to these geometries.
For a generic geometry with $f_\pm$  the Fourier modes of
$f_\pm$ are directly related to $L_{-n}$ and $\bar L_{-n}$ $n>1$ Virasoro charges \cite{BT,BGG-solution,GL-paper}.

To gain a better intuition about the geometries in \eqref{generic-solutions}, let us consider constant $f_\pm$ case, when both $x^\pm$ are periodic.
For \emph{constant (generic) non-zero and positive} $\bar f_\pm$ one recovers BTZ black hole. It is instructive to rewrite the metric into standard BTZ coordinates.
Under the coordinate transformation
\be
\label{coortran}
\rho^2=\frac{\left(r^2+\ell^2 \bar f_+\right) \left(r^2+\ell^2 \bar f_-\right) }{r^2},\quad t=\frac{\ell}{2}(x^++x^-), \quad \varphi=\frac{1}{2}(x^+-x^-)\,,
\ee
metric takes the form
\be
ds^2=-\frac{(\rho^2-\rho_+^2)(\rho^2-\rho_-^2)}{\ell^2\rho^2}dt^2+\frac{\ell^2\rho^2d\rho^2}{(\rho^2-\rho_+^2)(\rho^2-\rho_-^2)}+\rho^2\left(d\varphi-\frac{\rho_+\rho_-}{\ell \rho^2}dt\right)^2\,.
\ee
The BTZ inner and outer horizon radii $\rho_\pm$ and $\bar f_\pm$ are related as:
\be\label{BTZ-horizon-radii}
\bar{f}_\pm=\frac{1}{4\ell^2}(\rho_+\mp \rho_-)^2\,.
\ee
We note that the coordinate system \eqref{generic-solutions} is covering only the region of BTZ geometry which is outside the horizon.
As pointed out, Fourier modes of $f_\pm$ are related to charges of a Virasoro algebra which in turn is the symmetry of a 2d CFT dual to
the AdS$_3$ gravity theory \cite{BT, BGG-solution, GL-paper}. In the  standard AdS$_3$/CFT$_2$ terminology and for the constant $f_\pm$ case, then $\bar f_\pm$
are related to the energy of the Left and Right sectors of the dual 2d CFT $L_0,\ \bar L_0$ as \cite{AdS3/CFT2-reviews}
\be\label{L0-barL0-centralcharge}
L_0=\frac{\mathbf{c}}{6} \bar{f}_+\,,\qquad \bar L_0=\frac{\mathbf{c}}{6} \bar{f}_-\,,\qquad \mathbf{c}=\frac{3\ell}{2G_3}\,,
\ee
where $\mathbf{c}$ is the Brown-Henneaux central charge \cite{Brown-Henneaux} of the dual CFT.
Note also the positivity condition on $\bar f_\pm$.\footnote{These are generators of the Virasoro algebra for CFT on 2d
plane. For the 2d CFT on cylinder one should replace $L_0$ and $\bar L_0$ by $L_0+\mathbf{c}/24$ and $\bar L_0+\mathbf{c}/24$. The positivity condition on $\bar f_\pm$, which
is essentially the unitarity condition for the dual 2d CFT, then becomes $\bar f_\pm\geq -1/4$ \cite{AdS3/CFT2-reviews}.} The cases with negative but constant $f_\pm$, $ \bar f_\pm \leq -1/4$,
correspond to conic spaces, as we will see in the next section.

The class of solutions in \eqref{generic-solutions} have the following remarkable properties:
$\vspace{-0.1cm}$\bn
\item  are in one-to-one correspondence with the Brown-Henneaux diffeomorphisms;
\item for set of solutions with $f_\pm^i$, any linear combination/superposition of them specified by $f_\pm=\sum_i A_i f_\pm^i$ constitute another geometry within
the same class.
\item As we will show, for generic non-constant $f_\pm$ these geometries, while having (bifurcate) Killing horizon, the bifurcation curve is noncompact or not simply connected. Moreover, this Killing horizon is  not an event horizon.
    $\vspace{-0.1cm}$\en
Based on these facts, we propose to view the set of geometries with all possible $f_\pm$ which have the same average
\be\label{f0-average}
\bar{f}_\pm=\frac{1}{2\pi}\int_0^{2\pi}  f_\pm(x^\pm)\,,
\ee
as the ``fuzzball''\footnote{For discussion on the fuzzball proposal see \cite{fuzzball}.} \emph{semiclassical microstates} corresponding to the BTZ black hole
specified by $\bar{f}_\pm$. The mass and angular momentum of the BTZ black hole is given
through \eqref{L0-barL0-centralcharge},
\be
\ell M= L_0+\bar{L}_0  ,\quad J=L_0-\bar{L}_0.
\ee

In other words, a BTZ black hole with constant $\bar{f}_\pm$ may be viewed as ``classical geometries,''
and geometries in \eqref{generic-solutions} with a fixed average of $f_\pm$, as ``semiclassical'' description of BTZ geometries. The last step is to
``quantized'' geometries in \eqref{generic-solutions} to fully identify quantum microstates of a BTZ black hole. In this paper we intend to take the first step and establish
the geometries in \eqref{generic-solutions} as semiclassical microstates of BTZ black holes. The problem of quantization of these geometries is postponed to the next publication.

The rest of this paper is organized as follows. In sections \ref{sec-2} and \ref{sec-3}, we consider the geometries in which both $x^\pm$ are periodic and in section \ref{sec-4} consider geometries with periodic $x^-$ and while $x^+$ can be periodic or not. In section \ref{sec-2}, we analyze the structure of Killing vectors and isometries of metrics in \eqref{generic-solutions}.
We discuss local and global isometries and discuss the restrictions imposed on Killing vectors due to compactness of $x^\pm$. In section \ref{sec-3}, we study
the horizon structure of geometries in \eqref{generic-solutions}. We show that these geometries  generically have a bifurcate Killing horizon, however, the bifurcation surface (which is a one dimensional curve in the 3d case)  is not compact or simply-connected and is not in general an event horizon, except for
constant $f_\pm$ case or when one of $f_+$ or $f_-$ functions is vanishing. In section \ref{sec-4}, we repeat the analysis in sections \ref{sec-2} and \ref{sec-3} for the
case when $f_+=0$ and show that in this case the geometry has a degenerate (extremal) Killing horizon  \cite{LL-paper}, which is also an event horizon if $f_-(x^-)\geq 0$. These geometries may be viewed as
semiclassical fuzzball microstates of extremal BTZ black holes or AdS$_3$ self-dual orbifold. In section \ref{sec-5}, we summarize our results and present the outlook for the second part of this project, quantization of the semiclassical microstates. In three appendices we have gathered some technical details of computations. In the appendix \ref{Appen-A}, we
have shown that the algebra of the local Killing vectors is indeed $sl(2,R)\times sl(2,R)$.
This is done through defining a basic bracket among the functions upon which these Killing vectors fields have been constructed. In the appendix \ref{Appen-B},
 we have given brief review of mathematics literature on the solutions to Hill's equation, which appears
in our analysis of Killing vectors. In appendix \ref{LLvsUS}, we have gathered the analysis making a closer connection between our geometries with $f_+=0$ and those discussed in
\cite{LL-paper}.

\section{Local and global isometries of the general geometries}\label{sec-2}
To qualify as  semiclassical microstates we should study smoothness and  isometries and horizon structure of geometries in \eqref{generic-solutions}.
To this end we explore the Killing vectors of this geometry for any given $f_\pm(x^\pm)$. Note that in this section and also in section \ref{sec-3} we assume that
$f_\pm$ are both periodic functions of their argument, and that neither of them is zero (while some of the results in section \ref{section-2.1} is also true for more general cases, with noncompact $x^\pm$ or when $f_\pm$ vanish).

\subsection{Local isometries, the six Killing vectors}\label{section-2.1}

As discussed, these solutions are locally AdS$_3$ and  hence have  six Killing vectors. If we denote a given Killing vector by
\be
\zeta=h^1\partial_r+h^2\partial_{x^+}+h^3\partial_{x^-}\,,
\ee
then solving equation ${\mathcal L}_\zeta g_{\mu\nu}=0$ gives following relations
\bea\label{hi-killing}
h^1=-\frac{r}{2} ( K_+'+K_-'  ),\quad
h^2=K_+ + \frac{\ell^2r^2K_-''+\ell^4f_-K_+''}{2(r^4-\ell^4f_+f_-)},\quad h^3=K_- + \frac{\ell^2r^2K_+''+\ell^4f_+K_-''}{2(r^4-\ell^4f_+f_-)},
\eea
where $K_{\pm}=K_{\pm}(x^{\pm})$,  $prime$  denotes derivative with respect to the argument and functions $K_{\pm}$ satisfy following equations
\be\label{KV-conditions}
\begin{split}
K_+^{'''} - 4 K_+'f_+ - 2K_+f_+'&=0,\cr
K_-^{'''} - 4 K_-'f_- - 2K_-f_-'&=0\,.
\end{split}
\ee
Being two third-order differential equations, the above have $3\times 2$ independent solutions.
Our arguments below together with discussions in appendix \ref{Appen-A},
establish that, as expected, the geometries are locally invariant under $sl(2,R)\times sl(2,R)$. These isometries are also globally defined iff  \eqref{KV-conditions} have \emph{real and periodic} solutions.
The number of real periodic solutions to \eqref{KV-conditions} depends on the form of functions $f_\pm$ and can at most be six.

We remark that $h^a$ in \eqref{hi-killing}, in the leading order in large $r$ expansion,  have exactly the form of Brown-Henneaux diffeomorphisms \cite{Brown-Henneaux}, where in the latter $K_\pm=K_\pm(x^\pm)$ are two generic
functions, and are not subject to \eqref{KV-conditions}. Therefore, supposing that \eqref{KV-conditions} has at least a single real periodic solution for any
given $f_+$ and similarly for $f_-$, then there is a one-to-one correspondence between Brown-Henneaux diffeomorphisms and the geometries in \eqref{generic-solutions}. As we will explicitly show,
\eqref{KV-conditions} has indeed always two  real-periodic solutions (one corresponding to $K_+$ and one to $K_-$).
As stated in the introduction, our proposal is to identify the geometries with the same average $f_\pm$ and
with \emph{only one} real periodic $K_\pm$ as semiclassical microstates of the BTZ black hole.

\subsection{Constructing the Killing vector generators $K_a$}\label{section-2.2}

One can show that solutions to \eqref{KV-conditions} are specified through a Schr\"{o}dinger-type equation with periodic potential%
\be\label{Schrodinger}
\psi''-f\psi=0\,.
\ee
If the two linearly independent solutions to the above are denoted as $\psi_1$ and $\psi_2$, then it is easy to verify that
\be\label{Ka}
K_{-1}\equiv \frac12{\cal N}\psi_1^2\,,\qquad K_{+1}\equiv \frac12{\cal N}\psi_2^2\,,\qquad
K_{0}\equiv \frac12{\cal N}\psi_1\psi_2\,,\qquad
\ee
provide the three linearly independent solutions to \eqref{KV-conditions}. ${\cal N}$ is the normalization constant which is chosen to be
\be\label{normalization}
{\cal N}^{-1}=\psi'_1\psi_2-\psi_1\psi'_2\,.
\ee
(One can readily show that ${\cal N}$ is a constant of motion for \eqref{Schrodinger}.)
Hereafter, we will choose the normalization for $\psi_\alpha,\ \alpha=1,2$ such that ${\cal N}=1$. As has been shown in the appendix \ref{Appen-A}, based on
\eqref{normalization} one can define a Poisson bracket structure which explicitly shows that $K_a$'s indeed form an $sl(2,R)$ algebra. In other words, this Poisson bracket structure reproduces Lie bracket of the Killing vector fields given through $K_a$.

To qualify as Killing vectors we should also demand $K_a$'s to be real-valued.
Therefore, $\psi_\alpha$ are also real-valued. Periodicity of $K_a$'s is then related to periodicity of $\psi_\alpha$.
If $\psi_1$ and $\psi_2$ are both periodic then all three $K_a$'s are periodic too. However, even if $\psi_\alpha$ are not periodic, the Floquet theorem
(see Appendix B) implies that $\psi_1\psi_2$ is always periodic. Therefore, regardless of what $f_\pm$ are, we always have two compact global $U(1)_+\times U(1)_-$
isometries. This latter is the case associated with BTZ black hole semiclassical microstates.

To gain a better intuition about $\psi_\alpha$ and $K_a$, it is instructive to analyze  constant $f_\pm$ cases. One can distinguish two positive or negative $f$ cases:
\begin{itemize} \item If $f_\pm=\frac14 k_\pm^2>0$ then
\be\label{K's-BTZ}
\psi^\pm_1=e^{\frac12 k_\pm x^\pm}\,,\quad \psi_2=e^{-\frac12 k_\pm x^\pm}\,,
\ee
and hence the six \emph{local} Killings are generated by
\be\begin{split}
K_{\pm1}^+=\frac{1}{k_+}e^{\pm k_+x^+},\quad K_0^+=\frac{1}{k_+},\cr
K_{\pm1}^-=\frac{1}{k_-}e^{\pm k_-x^-},\quad K_0^-=\frac{1}{k_-}\,.
\end{split}
\ee
This case  corresponds to a BTZ geometry, where only two $K_0^\pm$  are periodic. This is of course compatible with our earlier discussions
where the BTZ geometry has two global $U(1)$'s while  is locally $sl(2,R)\times sl(2,R)$ invariant. Note that among the other four $K$'s,
two are exponentially growing and two and exponentially decaying with $x^\pm$.\\
\item For $f_\pm=-\frac14 k_\pm^2<0$,
\be\label{K's-conic}
\psi_1^\pm=\sin{\frac12 k_\pm x^\pm}\,,\quad \psi_2^\pm=\cos{\frac12 k_\pm x^\pm}\,,
\ee
and the six local Killings are generated by
\be\begin{split}
K_{\pm1}^+=\frac{1}{k_+}(1\pm\cos{k_+x^+})\,,\quad K_0^+=\frac{1}{k_+}\sin{k_+x^+}\,,\cr
K_{\pm1}^-=\frac{1}{k_-}(1\pm\cos{k_-x^-})\,,\quad K_0^-=\frac{1}{k_-}\sin{k_-x^-}\,.
\end{split}
\ee
For $f_\pm<0$ case, the Killings are not periodic for generic values of $k_\pm$.
The $k_\pm=1$ case is special because all six $K$'s are periodic with exactly period $2\pi$. This case with $f_\pm=-1/4$ corresponds to \emph{global} AdS$_3$ which
is the only case with  global $Sl(2,R)\times Sl(2,R)$ isometry group \cite{Deser-Jackiw}.

Integer $k_\pm$ cases are also special, because $K$'s are periodic for these cases too. One can interpret them as orbifolds of global AdS$_3$ by $Z_{{k_\pm}}$ with
the $Sl(2,R)/{Z_{k_+}}\times Sl(2,R)/{Z_{k_-}}$ global isometry. These geometries correspond to conic spaces and may be attributed to particles on AdS$_3$ \cite{Deser-Jackiw}.

\end{itemize}

\subsection{Further analysis of  global isometries of the generic geometries}

The Schr\"{o}dinger-type equation \eqref{Schrodinger} with a periodic potential is known as Hill's equation. Hill's equation have been extensively studied in
mathematics literature e.g. \cite{Hill-Eq-Book} (see Appendix \ref{Appen-B} for a brief review). Let the periodic function $f(x)$, $f(x)=f(x+2\pi)$,
have an average $\bar{f}$ (over $[0,2\pi]$ range) and a part $\hat{f}$ which is averaged to zero:
\be\label{f0-bar-f}
f(x)=\bar{f}+\hat{f}(x)\,,\qquad \bar{f}=\frac{1}{2\pi}\int_0^{2\pi}  f\,.
\ee
Then \eqref{Schrodinger} may be written as
\be\label{Schrod-2}
\psi''-\hat{f} \psi=\bar{f}\psi\,,
\ee
where $\hat{f}$ plays the role of a periodic potential oscillating around zero, and $-\bar{f}$ is the energy eigenvalues. Eq.\eqref{Schrod-2} has the form of standard Hill's equation
\eqref{Hill-Equation} with $\bar{f}=-\lambda$.

As already mentioned, Floquet theorem implies that $\psi_1\psi_2$ is always a periodic function. Moreover, since the operator $d^2/dx^2-\hat{f}$ is Hermitian,
$\psi_1\psi_2$ is also real-valued. Therefore, regardless of what $\hat{f}$ is, we have a global $U(1)_+\times U(1)_-$ isometry.

One may examine when all three $K_a$'s given in \eqref{Ka} are real-valued and periodic. This happens when both of $\psi_1$ and $\psi_2$ are real-valued and periodic.
This latter of course depends on the details of $\hat{f}$ and $\bar{f}$. This question has been addressed in the mathematics literature on the Hill's equation, which is reviewed in the
appendix \ref{Appen-B}. For a given $\hat{f}$ Hill's equation has periodic solutions for a discrete set of $\bar{f}$'s which are completely determined in
terms of $\hat{f}$. (Conversely, one can reconstruct $\hat{f}$ through its spectrum $\bar{f}$.) The values of $\bar{f}$ corresponding to periodic real solutions are arranged in two descending discrete sequences denoted by $\bar{f}_0,\bar{f}_1,\cdots$ and
$\bar{e}_1,\bar{e}_2,\cdots$, such that
\be\label{egienvalue-sequence}
\cdots \bar{f}_4\leq \bar{f}_3 <\bar{e}_4\leq \bar{e}_3 <\bar{f}_2 \leq\bar{f}_1<\bar{e}_2\leq \bar{e}_1 <\bar{f}_0\,,
\ee
and the largest one $\bar{f}_0$ has a positive value bounded as
\be\label{f0-bound}
0\leq \bar{f}_0\leq \frac{1}{2\pi}\int_0^{2\pi} \hat{f}^2\equiv \overline{\hat{f}^2}\,.
\ee
There is some value of $n$ where $\bar{f}_n\ (n\geq 1)$ changes sign and becomes negative.
For large values of $k$ and smooth functions $f_\pm$, $\bar{f}_k$ and $\bar{e}_{k}$ respectively approach to $-k^2$ and $-(k-1/2)^2$ (with ${\cal O}(1/k^2)$ corrections,
\emph{cf.} \eqref{lambda-mu-large-k}).  This is remarkable, recalling the constant, negative $f_\pm$ case discussed in the previous subsection: For large $k$, regardless of the details of $f_\pm$,
we will have a geometry whose Killing vectors approach the conic space.
In other words, for a given $\hat{f}$ and $\bar{f}$ taking negative large values, the geometries in \eqref{generic-solutions} may be viewed as ``semiclassical bound state'' of conic space (particles on AdS$_3$) and Brown-Henneaux boundary gravitons.\footnote{If $f$ is such that $\bar{f}$ exactly coincides with one of the $\bar{f}_k$'s (with $f_k<-1/4$) then the geometry is generically just a conic space.}

For given geometries specified through $f_\pm(x^\pm)$, however,  $\bar{f}_\pm=\frac{1}{2\pi}\int_0^{2\pi} f_\pm$ does not generically coincide with
any $\bar{f}_k$'s  associated with functions $\hat{f}_\pm(x^\pm)$. So, we do not generically  have three periodic
$K_a$'s. In this case, $\bar{f}$ may fall into stability or into instability intervals. If $\bar f$ is in
the \emph{stability intervals} of Hill's equation, solutions  $\psi$'s are not periodic, nonetheless they fluctuate/oscillate  around a periodic solution.
These geometries may hence be viewed as ``excitations'' around AdS$_3$ particles (conic spaces).
On the other hand if $\bar f$ falls into \emph{instability intervals}, in particular if $\bar{f}>\bar{f}_0$, one of the solutions has exponential growth while the other is exponentially suppressed and we have a geometry which may be viewed
as ``excitations'' around a BTZ black hole.
In fact these are the geometries which we would like to identify as ``semiclassical microstates'' of the BTZ black hole.

\section{Horizon structure of the general geometries}\label{sec-3}

Killing horizon is a codimension one null hyper-surface where norm of a Killing vector vanishes.
There are six Killing vectors formed by linear combination of functions $K^\pm_0$ and $K_{\pm1}^\pm$ which are bi-linears made out of
solutions to Schr\"{o}dinger equations $\psi''-f_+\psi=0$ and $\phi''-f_-\phi=0$. Out of these six, two of them generated by $K^\pm_0$ are periodic and globally defined. If we denote these two normalized solutions as $\psi_\alpha$ and $\phi_\alpha$,
one can show that the norm of the most general combination of the two globally defined Killing vectors may always be brought to the form
\be\label{most-general-Killing-horizon}
|\zeta_H|^2=-\frac{{\mathcal K}^2}{r^2}\biggl(r^2\psi_1(x^+)\phi_1(x^-)-\ell^2\psi_1'(x^+)\phi_1'(x^-)\biggr)
\biggl(r^2\psi_2(x^+)\phi_2(x^-)-\ell^2\psi_2'(x^+)\phi_2'(x^-)\biggr)\,,
\ee
where
\be\label{Killing-horizon-zeta}
\zeta_H=h^1\partial_r+h^2\partial_{x^+}+h^3\partial_{x^-}
\ee
with $h^i$ given in \eqref{hi-killing} and
\be\label{K-Killing-horizon}
K_+={\cal K}\psi_1(x^+)\psi_2(x^+)\,,\qquad K_-={\cal K}\phi_1(x^-)\phi_2(x^-)\,,
\ee
and ${\cal K}$ is an arbitrary constant, which as we will show below, is equal to the surface gravity by an appropriate choice of normalization for $\zeta_H$.

\paragraph{Killing horizons.} Geometries in \eqref{generic-solutions} have a Killing horizon at the locus $\zeta_H$ vanishes. This happens at ${\cal H}(r,x^+,x^-)=0$ where
\be\label{Killing-horizon}
{\cal H}=r^2\psi_1(x^+)\phi_1(x^-)-\ell^2\psi_1'(x^+)\phi_1'(x^-)\,.
\ee
One can then readily show that
\be\label{nabla-N-sq}
(\nabla {\cal H})^2=\frac{4r^2}{\ell^2} \psi_1(x^+)\phi_1(x^-)\cdot {\cal H}\,.
\ee
Therefore, ${\cal H}=0$ is a null surface. The Killing horizon is located on the two dimensional surface
\be\label{rH}
r^2=r_H^2=\ell^2\frac{\psi_1'(x^+)}{\psi_1(x^+)}\frac{\phi_1'(x^-)}{\phi_1(x^-)}\,.
\ee
Note that \eqref{most-general-Killing-horizon} has another solution, $\tilde r_H^2$,
\be\label{tilde-rH}
\tilde r_H^2=\ell^2 \frac{\psi_2'(x^+)}{\psi_2(x^+)}\frac{\phi_2'(x^-)}{\phi_2(x^-)}\,.
\ee
We will be assuming that $r_H$ is the outer horizon, i.e. $r^2_H\geq \tilde r_H^2$ for any given $x^\pm$.

\paragraph{Surface gravity.} Given Killing vector field \eqref{Killing-horizon-zeta} one can compute the surface gravity $\kappa$:
\be\label{surface-gravity-def}
\zeta^\mu\nabla_\mu \zeta_\nu\big|_{{\cal H}=0}=\kappa \zeta_\nu\big|_{{\cal H}=0}\,.
\ee
Performing the straightforward analysis shows that $\kappa={\cal K}$, which, as expected, is a constant. We note that, as the BTZ example below indicates,
the value of surface gravity $\kappa$ is specified by the $\bar f_\pm$ (through normalized wavefunctions $\psi$ and $\phi$).

Since the surface gravity is non-zero, the Killing horizon, if exists, is a bifurcate horizon. The cases with vanishing surface gravity, the extremal horizon,
is a different family which will be discussed in section \ref{sec-4}.

\paragraph{BTZ example.} When $f_+$ and $f_-$ are positive constants, $K_\pm$ are also constant. With our normalization $K_\pm=\kappa/(2\sqrt{f_\pm})$, and horizon is at
${\cal H}_{BTZ}=0$ where,
\be
{\cal H}_{BTZ}=e^{\sqrt{f_+}x^++\sqrt{f_-}x^-}(r^2-\ell^2\sqrt{f_+f_-})\,.
\ee
The standard BTZ normalization for the $\zeta_H$ is such that, $\zeta_H=\partial_t +\Omega\partial_\varphi$ where
\be\label{kappa-Omega-BTZ}
\kappa=\frac{2\sqrt{f_+f_-}}{\sqrt{f_+}+\sqrt{f_-}}\,,\qquad \Omega=\frac{\sqrt{f_+}-\sqrt{f_-}}{\sqrt{f_+}+\sqrt{f_-}}.
\ee
In the extremal BTZ case, $f_-=0$ (or $f_+=0)$ and it is readily seen that surface gravity vanishes, $\kappa=0$, and $\Omega=1$ (or $\Omega=-1$).

\subsection{Is the bifurcate Killing horizon compact and simply-connected?}

As mentioned  the Killing horizon, which we showed it always exists, is generically a bifurcate horizon. We then have a ``bifurcation curve'' which is a one dimensional curve on the $r^2=r^2_H$ surface. The vectors $\zeta_H$ and $\nabla {\cal H}$ should be normal to this curve. Therefore, we can already find the vector along (tangent to) the bifurcation curve.
One can show that there is another linear combination of globally defined Killing vectors generated by $K_0^+$ and $K_0^-$, denoted  by $T$, which is tangent to the bifurcation curve. This Killing vector field, as in the BTZ case, remains spacelike everywhere outside horizon, as long as $|\zeta_H|^2\leq0$. A straightforward computation shows that
$|T|^2=\ell^2$ at the horizon.\footnote{The fact that $\zeta_H$ remains time-like everywhere outside horizon is particular to BTZ (and possibly other asymptotic AdS) black holes. For the
generic stationary asymptotic black holes, like e.g. Kerr-Neumann black hole, $\zeta_H$ is asymptotically space-like.}
The Killing vector $T$ is tangent to the bifurcation curve  and the necessary (but not sufficient) condition for having an \textit{event horizon} is $T$ to be periodic. Periodicity of $T$ is guaranteed iff $K_0^\pm$ are periodic. This latter follows from the Floquet theorem.

To explore presence of an event horizon, we first analyze compactness and simply-connectedness of the possible Killing horizon. In what follows we have enumerated the conditions for having a compact or simply-connected bifurcation curve:

\bn\item ${\cal H}=0$ \eqref{Killing-horizon} has acceptable solutions. That is, we need to require
\be\label{KH-1}
r_H^2=\ell^2\frac{\psi_1'(x^+)}{\psi_1(x^+)}\frac{\phi'_1(x^-)}{\phi_1(x^-)}\geq 0\,,
\ee
everywhere on the $x^\pm$ plane.\\
\item 
Horizon radius $r_H^2$ should always remain larger than $\tilde{r}_H^2$ \eqref{tilde-rH}. That is,
\be\label{KH-2}
\delta r_H^2\equiv \frac{\psi_1'(x^+)}{\psi_1(x^+)}\frac{\phi_1'(x^-)}{\phi_1(x^-)}-\frac{\psi_2'(x^+)}{\psi_2(x^+)}\frac{\phi_2'(x^-)}{\phi_2(x^-)}\,,
\ee
should not change sign on $x^\pm$ plane.
\item Generators of $K_0^\pm$ and also $r^2_H$ should be periodic functions on $x^\pm$ plane, i.e.
\be\label{periodic-K-r2}
\begin{split}
&\frac{\psi_1'}{\psi_1}\ is\ periodic,\qquad  {\psi_1}{\psi_2}\ is\ periodic\,,\cr
&\frac{\phi_1'}{\phi_1}\ is\ periodic,\qquad  {\phi_1}{\phi_2}\ is\ periodic\,.
\end{split}
\ee
These conditions are fulfilled, thanks to the Floquet theorem (see Appendix \ref{Appen-B}).\\

\item Bifurcation curve should be compact or simply-connected. Suppose that the bifurcate curve is  given by
\be
r=r(s)\,,\quad x^+=x^+(s)\,,\quad x^-=x^-(s)\,,
\ee
where $s$ parameterizes the one dimensional bifurcation curve. We choose the range of $s$ such that $s\in[0,s_0]$. As discussed above, the tangent to this curve $T$ is along a linear combination of $K_0^\pm$ Killings (which is a space-like Killing vector at $r=r_H$ given in \eqref{rH}). Explicitly,
\be
\frac{dr}{ds}=\xi^r\,,\quad \frac{dx^+}{ds}=\xi^+\,,\quad  \frac{dx^-}{ds}=\xi^-\,,
\ee
where the derivatives are computed on the horizon and $\xi^\mu$ are components of $T$ and given in \eqref{hi-killing} with
$$
K_0^+=A\psi_1\psi_2\,,\qquad K_0^-=B\phi_1\phi_2\,,
$$
where $A, B$ ($A\neq B$) are two constants such that $T\cdot\nabla {\cal H}|_{r=r_H}=T\cdot \zeta_H|_{r=r_H}=0$. Length of horizon is then given as
$$
L=\int_0^{s_0} |T|^2 ds=\ell\int_0^{s_0} ds=\ell s_0\,.
$$
Therefore, the compactness condition of the bifurcate curve translates into the finiteness of $s_0$. This condition and the connectedness condition of bifurcation curve is satisfied if $\xi^\mu$
are finite everywhere at the horizon. This means that the denominator in \eqref{hi-killing} should not vanish at the horizon, or
\be\label{r4-condition}
r_H^4>\ell^4 f_+ f_-\,.
\ee
There are, however, two important exceptions to the above condition:\vskip 1mm
\bn\item $r_H^4=\ell^4 f_+f_-$ but $K''_\pm=0$ so that the numerator in \eqref{hi-killing} also vanishes at $r_H$. This latter can only happen for the constant
$f_\pm$, the BTZ case.\vskip 1mm
\item When $f_-$ or $f_+$ is vanishing. This latter is the case discussed in \cite{LL-paper} and we will come to it in the next section.
\en
So, hereafter we exclude the above two cases and explore  nonconstant, nonvanishing $f_\pm$ cases.
\en
\vskip 3mm
Positivity of $r^2_H$, noting that $\psi$ and $\phi$ are completely independent functions, imply that
\be\label{event-horizon-cond.}
\frac{\psi'}{\psi}\,, \frac{\phi'}{\phi} \ are\  periodic\ continuous\ and\ have\ a\ definite\ sign.
\ee
On the other hand, recalling that $\psi$ and $\phi$ are both solutions to Schr\"{o}dinger equation with periodic potentials $f_\pm$
$$
\psi''-f_+\psi=0\,,\qquad \phi''-f_-\phi=0,
$$
if both $f_\pm$ are nonvanishing, \eqref{r4-condition} may be written as
$$
\left(\frac{\psi'}{\psi}\right)^2\left(\frac{\phi'}{\phi}\right)^2> \frac{\psi''}{\psi}\frac{\phi''}{\phi}\,,
$$
which is satisfied iff
\be
\left(\frac{\psi'}{\psi}\right)^2>\frac{\psi''}{\psi},\quad \text{or}\quad \left(\frac{\phi'}{\phi}\right)^2>\frac{\phi''}{\phi}\,,
\ee
where we used the fact that $\psi$ and $\phi$ are two independent functions of different arguments.
The above may also be written as
\be\label{r4-condition-processed}
\left(\frac{\psi'}{\psi}\right)'<0,\quad \text{or}\quad \left(\frac{\phi'}{\phi}\right)'<0\,.
\ee
Nonetheless, \eqref{r4-condition-processed} and \eqref{event-horizon-cond.} cannot be simultaneously satisfied because a periodic continuous function should necessarily have points
where its derivative changes sign. We hence conclude that the only cases with compact and simply-connected Killing horizon are exactly the two cases excluded above, that is, the constant $f_\pm$ (the BTZ) case or the degenerate horizon case with $f_-$ or $f_+$ equal to zero. This is accord with the discussions of \cite{LL-paper}.

\subsection{Is the Killing horizon also an event horizon?}\label{section-3.2}

So far we showed that geometries in \eqref{generic-solutions} generically have a bifurcate Killing horizon and argued that  the bifurcate curve  is generically noncompact or not simply-connected. In order to qualify as a black hole, however, it is necessary to have event horizon. This is more crucial for the AdS$_3$ black holes, as all the solutions to AdS$_3$ Einstein gravity are locally AdS$_3$. Unlike the Killing horizon which is a local feature of the geometry, the notion of event horizon is a global property of the geometry, e.g. see \cite{Black-hole-review} for more detailed discussions.

For asymptotic AdS spacetimes, like in our case, future (or past) event horizon $\mathcal{H}_+$ (or $\mathcal{H}_-$) is defined as the boundary of the closure of causal past (or future) of the points on the boundary of the AdS space.  The Penrose theorem (see e.g. \cite{Black-hole-review}) stating that the generators of $\mathcal{H}_+$ (or $\mathcal{H}_-$) have no future (or past) endpoints, also extends to the asymptotic AdS case. In terms of the causal structure and the Penrose diagram this means that the past or future event horizon must only intersect the causal boundary in two points (once in the ``infinite past'' and once in the ``infinite future''). If the horizon does not intersect the boundary, this means that one can circumvent the horizon and pass to the other side at finite AdS coordinate time and hence we do not have a black hole.

Intuitively, having an infinite length noncompact (or not simply-connected) bifurcation curve on the Killing horizon may seem to be suggesting that it either does not intersect the AdS boundary or it intersects it many times (in one period of $x^\pm$) and hence, is not an event horizon. In this part we bring arguments
that the Killing horizon with noncompact (or not simply-connected) bifurcation curve is not an event horizon.
To this end, let us focus on the null geodesics of metric \eqref{generic-solutions}. Since we are interested in the past (or future) null geodesics ending at the causal boundary of asymptotic AdS, it is more convenient to adopt the coordinate system which makes manifest the time and space part of the 2d boundary surface. These coordinates $t,\varphi$ were introduced in \eqref{coortran}
in which metric (\ref{generic-solutions}) takes the following form
\be\label{metric-t-phi}
ds^2 =\frac{-(r^4-\ell^4f_+f_-)^2 dt^2 }{\ell^2r^2(r^2+\ell^2f_+)(r^2+\ell^2f_-)}+\frac{\ell^2dr^2}{r^2}
+\hat\rho^2\left(d\varphi +\frac{\ell(f_+-f_-)dt}{\hat\rho^2}\right)^2,
\ee
where
$$
\hat\rho^2=\frac{(r^2+\ell^2f_+)(r^2+\ell^2f_-)}{r^2}\,.
$$
In this coordinate system  equation of the radial (zero angular momentum) null geodesic is
\be\label{Radial-Null-geod}
\dot{r}^2=\frac{(r^4-\ell^4f_+f_-)^2}{\ell^4(r^2+\ell^2f_+)(r^2+\ell^2f_-)} \quad \Longrightarrow\quad \dot r=\pm \frac{1}{\ell^2 r\hat\rho} (r^4-\ell^4f_+f_-)
\ee
where $dot$ denotes derivative with respect to $t$. The plus (minus) sign corresponds to future (past) oriented geodesic.

For our analysis we should focus on the past oriented null geodesics starting from an arbitrary point at the AdS boundary and examine if it reaches and intersects the horizon or not. For the ``radial null geodesics'' described in \eqref{Radial-Null-geod}, since $\dot r$ changes sign at $r_0^4\equiv\ell^4f_+f_-$, for having an event horizon we need to require that $r_H^4\geq r_0^4$, or equivalently the necessary condition for having an event horizon is essentially the same condition for compactness and connectedness of the horizon \eqref{r4-condition}. (The special case of $r_H^4=\ell^4f_+ f_-$ corresponds to either of the two ``exceptional cases'' discussed in previous subsection, for which we do have event horizon.)

In view of the argument above, few comments are in order:
\bi
\item Our discussions above can be readily  extended to the  more general non-radial null geodesics, noting that addition of angular momentum will add a positive-definite  contribution to the left-hand-side of the first equation in \eqref{Radial-Null-geod}.
\item The condition for having event horizon and compact and connectedness of the bifurcation curve on the Killing horizon has come about because $f_\pm$
generically  have $x^\pm$ dependence and this makes the Killing horizon radius $r_H$ \eqref{rH} be different than $r_0\equiv \ell(f_+f_-)^{1/4}$.
\item $r_0$ is basically the end of the range for the coordinate system used in \eqref{generic-solutions}; in particular we note that $\sqrt{-\det g}=\frac{\ell}{2r^3}(r^4-\ell^4f_+ f_-)$.
\footnote{Here we have shown that there is no Killing horizon which is also an event horizon and there is still the possibility of having a compact and simply-connected event horizon which is not Killing. To rule out the latter and complete proof of nonexistence of event horizons, the above analysis should be supplemented by a full analysis of the causal structure of the geometries \eqref{generic-solutions} which may extend coordinates beyond $r_0$. This latter will be postponed to future publications.}
\ei

The above argument completes our prerequisites enumerated in  the introduction for viewing generic geometries \eqref{generic-solutions} as semiclassical fuzzball microstates of BTZ black holes. This will be discussed further in section \ref{sec-5}.

\section{The degenerate horizon case}\label{sec-4}
In the previous section we showed that when $f_\pm$ functions in metric  \eqref{generic-solutions} are non-zero and are not constants, the geometry has a bifurcate Killing horizon
but not an event horizon. Our argument for nonexistence of event horizon, however, had two exceptions, constant $f_\pm$ and when one of
$f_-$ or $f_+$ is vanishing. The constant $f_\pm$ case as discussed corresponds to BTZ black hole which admits event horizon \cite{BTZ}.
In this section we discuss the other case, by setting $f_+=0$, and consider metric of the form
\be\label{f+=0-metric}
ds^2=\frac{\ell^2 dr^2}{r^2}-dx^-(r^2dx^+-\ell^2f_-dx^-)\,.
\ee

We start analysis of metric \eqref{f+=0-metric} by studying its Killing vector fields.
The Killing vectors are given by \eqref{hi-killing} but now, $K_+$ and $K_-$ satisfy
\be\label{K+K-Extremal}
K_+''' =0\,,\qquad  K_-'''-4K_-'f_--2K_-f_-'=0\,.
\ee
The most general solutions for the $K_+$ equation is a linear combination of the three solutions
\be\label{K+f+=0}
K_{+,-1}=1\,,\quad K_{+,0}=x^+\,,\quad K_{+,+1}=(x^+)^2\,.
\ee
The Killing vectors associated with these three solutions form an $sl(2,R)$ algebra. It is explicitly seen that only the generator corresponding to $K_{+,-1}$ (this is generator of
translations along $x^+$ direction) is periodic in $x^+$.\footnote{One may represent the three $K$'s in \eqref{K+f+=0} in terms of solutions to Schr\"{o}dinger-type
 equation $\psi''=0$, as in \eqref{Ka}. The solutions of this equation are $\psi_1=1,\ \psi_2=x^+$, and hence $K_{+,-1}=\psi_1^2,\ K_{+,0}=\psi_1\psi_2,\
 K_{+,+1}=\psi_2^2$. Note that, unlike the generic case with non-zero $f_+$,  in this case the $K_{+,-1}$, and not $K_{+,0}$, is periodic; for this special case Floquet theorem does not apply.} It is readily seen that the norm of this global Killing vector field, $\partial_{x^+}$, is zero and
hence the geometry has a globally defined null Killing vector field and a Killing horizon at $r=0$. This geometry should hence fall into the class constructed and discussed in \cite{LL-paper}.
One can indeed construct a coordinate transformation which (locally) maps the geometry \eqref{f+=0-metric} to the one in \cite{LL-paper} (through this map, the function
$f_-(x^-)$ will be mapped onto the function $\beta(x)$, see eq.(31) of \cite{LL-paper}). More detailed analysis of the Killing algebra of the geometries in \cite{LL-paper}
and their connection to the coordinate system we have adopted in this paper is gathered in the appendix \ref{LLvsUS}.

Depending on the compactness of $x^+$, we have two distinct classes of geometries:
\bn
\item Compact $x^+$ case, where \eqref{f+=0-metric} corresponds to geometries which are either ``semiclassical microstates'' (or excitations) of an extremal BTZ black hole
when $\bar f_-\geq 0$ or to an ``extremal'' conic space (an AdS$_3$ particle with equal mass and angular momentum) when $\bar f_-< 0$. The special case of $f_-=0$ corresponds to massless BTZ.

The discussions
in this case is essentially the same as the generic case of previous two sections, with the important difference that we have degenerate (extremal)
Killing horizon. The metric at Killing horizon which is located at $r=0$ is
\be\label{metric-degerate-horizon}
ds^2_H=f_-\cdot (dx^-)^2\,.
\ee
Therefore, we have an event horizon if $f_-$ is positive function over its range $x\in [0,2\pi]$.
The ``area of horizon'' is then $ s_0=\ell\int_0^{2\pi} \sqrt{f_-} dx^-$.
For constant  $f_-$ case, this matches with the standard extremal BTZ result, $s_0=2\pi \ell \sqrt{f_-}$.

\vskip 3mm
\item Noncompact $x^+$ case, while $x^-$ is still compact. In this case and for generic nonconstant $f_-$ the geometry has a global $R\times U(1)$ isometry.
In the constant, positive $f_-$ case, this geometry corresponds to the Self-Dual AdS$_3$ Orbifold (SDO) and has a global $Sl(2,R)\times U(1)$ isometry \cite{BTZ-SDA}.
This geometry does not have an event horizon while it has a Killing horizon at $r=0$. The two compact and noncompact $x^+$ cases are related
    through the \emph{near-horizon} limit:
\be\label{NH-limit}
r=\epsilon\rho\,,\qquad x^+=u/\epsilon^2\,
\ee
while keeping $x^-,\ u$ and $\rho$ fixed.  As was discussed in \cite{DLCQ-paper}, the above scaling decompactifies $x^+$. On the dual 2d CFT this near horizon limit corresponds to performing a
Discrete Light-Cone Quantization (DLCQ) \cite{DLCQ-paper}.
\en
Since the two cases are related by the near-horizon limit \eqref{NH-limit} which keeps the $x^-$ dependent part of the metric \eqref{f+=0-metric} intact,
we may discuss both cases simultaneously. These geometries have three other Killing vectors (local isometries) associated with three solutions of \eqref{K+K-Extremal} for $K_-$.
These three Killing vectors are specified through the solutions to the Schr\"odinger-type equation
\be\label{phi-Schrod}
\phi''-f_-\phi=0\,,
\ee
which as discussed in subsections \ref{section-2.1} and \ref{section-2.2}, form a local $sl(2,R)$ algebra.
The Floquet theorem again implies that the generator of the $u(1)\in sl(2,R)$, the $K_0$, is periodic in $x^-$. This result is in
agreement with the analysis in \cite{LL-paper}. For more details see appendix \ref{LLvsUS}.
This $sl(2,R)$ can become a global one only if $f_-=-1/4$ and is a constant.
Then, depending on the average value of $f_-$, $\bar{f}_-$, \eqref{phi-Schrod} may have periodic, stable or unstable solutions (see appendix \ref{Appen-B}).

Our proposal is specialized for these cases as follows: Geometries in \eqref{f+=0-metric} with generic positive $f_-(x^-)$, are semiclassical microstates of
extremal black holes/SDO specified by the average of $f_-(x^-)$ function, $\bar f_-$. These semiclassical microstates are then labeled by the Fourier modes of
function $f_-$, which are nothing but the conserved charges associated with a chiral-half of a Virasoro algebra generated by the Brown-Henneaux
diffeomorphisms \cite{LL-paper,{GL-paper},DLCQ-paper}.\footnote{We note that, as discussed, these geometries have degenerate event horizon. In the dual CFT, these geometries correspond to those appearing in a zero temperature ``thermal state.'' These states constitute the ground state of a chiral sector of the 2d CFT, or equivalently,  the vacuum state for the DLCQ of the dual 2d CFT \cite{DLCQ-paper}. This is to be compared with the generic non-degenerate horizon case where the geometries without event horizon are proposed to describe excitations above the 2d CFT vacuum state (which corresponds to massless BTZ).   }

\section{Concluding remarks and outlook}\label{sec-5}
In this work we revisited the problem of AdS$_3$ gravity and took the first step toward its quantization. Our starting point was recalling that all solutions to
AdS$_3$ Einstein gravity are locally AdS$_3$ and that these geometries can be classified according to their global features (like topology of the asymptotic geometry)
and more importantly, by the behavior of the metric in the asymptotic AdS$_3$ region. The most general geometries  with the Brown-Henneaux boundary behavior are
geometries in \eqref{generic-solutions}. These solutions are specified by two ``holomorphic'' functions $f_+(x^+)$ and $f_-(x^-)$ which in the standard AdS$_3$/CFT$_2$ dictionary \cite{BGG-solution,Sol-Sken,{AdS3/CFT2-reviews}}
are related to the energy momentum of the dual 2d CFT.

Here, we used the fact that all the geometries in \eqref{generic-solutions} are diffeomorphic to AdS$_3$ (up to ``large'' coordinate transformations).
Therefore, all the information in these geometries are equally encoded in the large diffeomorphisms which map them to AdS$_3$. This viewpoint is complementary to
that of standard the Brown and Henneaux \cite{Brown-Henneaux}, where diffeomorphisms with specific boundary conditions, which take us away from the AdS$_3$, are considered and analyzed.
The two viewpoints are of course closely related as the comparison between set of our Killing vectors \eqref{hi-killing} and the Brown-Henneaux diffeomorphisms
indicates. In other words, Brown-Henneaux boundary gravitons are in one-to-one relation with the set of Killing vectors associated to a geometry with given $f_\pm$.
The diffeomorphisms which map our geometries to AdS$_3$ are completely specified by the form of the Killing vectors of the geometries. Being locally AdS$_3$,
there are  six such Killing vector fields which form local $sl(2,R)\times sl(2,R)$ isometry algebra.  So, we focused on studying these Killing vectors.

Further analysis of the geometries in \eqref{generic-solutions} and their Killing vectors fields revealed a very interesting feature: Generic geometries in this class  have always  (bifurcate)
Killing horizon, which is of course generated by a linear combination of these Killings and is a local feature of spacetime, but do not generically have a compact (finite length) and simply-connected bifurcation curve. This latter is compatible with the analysis of \cite{LL-paper}. Moreover, we  argued that these Killing horizons are not \emph{event horizons}. Our analysis for the latter was presented in section \ref{section-3.2}, but a more thorough analysis is needed to prove that these geometries do not have compact event horizons which are not Killing horizons.
This latter is an unlikely possibility if we restrict to geometries which satisfy Brown-Henneaux  boundary conditions, as we did, noting that these geometries are specified by the conserved charges uniquely determined through functions $f_\pm$. Further analysis on this issue is  postponed to future works.\footnote{We thank James Lucietti and Steve Carlip for discussions on this issue.}
Stationary black holes,  as BTZ black holes, on the other hand, are geometries with event horizon. So, as we discussed in detail
our proposal is to view the geometries in \eqref{generic-solutions} which  are not generically black holes, as ``semiclassical'' microstates around a BTZ black hole
 specified by the average of functions $f_\pm$ (\emph{cf} \eqref{f0-average}). This is a direct realization of the fuzzball proposal \cite{fuzzball}:
Each of these semiclassical microstates, the fuzzball geometries, are specified by two functions $f_\pm$ or equivalently by their Fourier modes and are
superposed/averaged to a BTZ black hole.

This picture is very simple and interesting, but the number of these microstates are not large enough to account for the entropy of the associated BTZ black hole. We need
to perform quantization over these semiclassical microstates to get a full quantum description of BTZ black hole microstates and probably then an AdS$_3$ quantum gravity.\footnote{See the papers by S. Carlip, e.g. \cite{Carlip}, for related analysis and ideas.}
Quantization of these geometries is what will be studied in our next publication. However, here, we make some remarks on our vision for this quantization.
One may follow either of the two ideas outlined below:
\paragraph{1. Promoting $f_\pm$ to operators $\boldsymbol{f_\pm}$}such that their ensemble average equals their space average
\be\label{quantization-VEV}
\bar f_\pm=\frac{1}{2\pi}\int_0^{2\pi} f_\pm= \langle \boldsymbol{f}_\pm\rangle.
\ee
This quantization proposal is closely related to the one in standard Brown-Henneaux analysis \cite{BT,BGG-solution,Brown-Henneaux,Banados,GL-paper}, where
one is prescribed to directly relate Fourier modes of $f_\pm$ to the Virasoro algebra generators, or equivalently,
\be\label{f-quantized}
\boldsymbol{f}_+=\frac{6}{\mathbf{c}} \sum_{n\in \mathbb{Z}}\ \mathbf{L}_n e^{in x^+}\,,\qquad \boldsymbol{f}_-= \frac{6}{\mathbf{c}}\sum_{n\in \mathbb{Z}}\
\mathbf{\bar{L}}_n e^{in x^-}\,,
\ee
where $\mathbf{L}_n,\ \mathbf{\bar{L}}_n$ are Virasoro generators,
\be\label{Virasoro-algebra}
[\mathbf{L}_n,{\mathbf{L}_m}]=(m-n)\mathbf{L}_{m+n}+\frac{\mathbf{c}}{12}n(n^2-1)\delta_{m+n}\,,\qquad
[\mathbf{\bar{L}}_n,\mathbf{\bar{L}}_m]=(m-n)\mathbf{\bar{L}}_{m+n}+\frac{\mathbf{c}}{12}n(n^2-1)\delta_{m+n}\,,
\ee
and $\mathbf{c}$ is the Brown-Henneaux central charge \eqref{L0-barL0-centralcharge}. This quantization proposal is basically the standard  Weyl correspondence rule of quantum mechanics applied to AdS$_3$ gravity within our proposal.
With the above choice and for any state in the Hilbert space which form a
representation of the above Virasoro, condition \eqref{quantization-VEV}
is automatically satisfied. One should, however, note that these ``semiclassical'' microstate geometries correspond to ``mixed states'' (rather than pure states) in the
Virasoro representations or in the corresponding dual 2d CFT.

\paragraph{2. Boundary gravitons as ``edge-states.''} Another idea for quantization, which is of course closely related to the previous idea, is to
view Brown-Henneaux boundary gravitons as edge-states in a quantum Hall system. This idea is based on the observation that AdS$_3$ Einstein gravity,
at least classically,  is equivalent to
an $sl(2,R)\times sl(2,R)$ Chern-Simons gauge theory \cite{3d-gravity,AdS3-gravity-Witten}:
\be\label{CS-gravity}
S_{\text{CS}}=\frac{1}{16\pi G}\int d^3x \sqrt{-g} (R-\frac{1}{\ell^2})=\frac{\kappa}{4\pi}\int d^3x \left({\cal L}_{CS}(A^+)-{\cal L}_{CS}(A^-)\right)\,,
\ee
where
\be
{\cal L}_{CS}(A)=\mathrm{Tr} \epsilon^{\mu\nu\alpha}\left( A_\mu\partial_\nu A_\alpha+\frac{2i}{3} A_\mu A_\nu A_\alpha\right)
\ee
and $A^\pm_\mu$ are $sl(2,R)$ valued gauge fields, which are components of the one-forms
\be
A_\pm^a=\mathbf{\omega}^a\pm \frac{1}{\ell} \mathbf{e}^a\,,
\ee
with $\mathbf{\omega}^a$ being the $sl(2,R)$ spin connections and $\mathbf{e}^a$ the dreibein, and $\kappa=\frac{\ell}{4G}=\frac{\boldsymbol{c}}{6}$.

On the other hand, quantum Hall system may also be described by a Chern-Simons theory \cite{Susskind-QHS}. Chern-Simons theory, and hence the corresponding
quantum Hall system is a topological theory and hence the ``dynamics'' of system is completely determined by the ``edge states''
describing the boundaries of the quantum Hall droplets, e.g. see \cite{Edgestate-QHY} and reference therein. The idea we would like to put forward is somehow paralleling the same
idea as edge-states: The quantum dynamics of AdS$_3$ gravity is described by its ``edge states''. In a sense this idea is not new, as notion of Brown-Henneaux boundary gravitons
is essentially very similar.\footnote{It is also inspiring to recall that gravity on AdS space, due to the causal boundary of AdS, is basically behaving like  gravity in a box. This box
may be thought as the quantum Hall droplet in our analogy. Then the AdS boundary, is the edge of the droplet.}$^,$\footnote{See e.g. \cite{Edge-state-references} for some related discussions.}
However, what makes this idea more tractable is the detailed
analysis of the Killing vectors of most general semiclassical geometries associated with the Brown-Henneaux boundary gravitons, i.e. geometries in \eqref{generic-solutions}.
As already mentioned, there is a one-to-one correspondence between the Killing vectors of these geometries and boundary gravitons. In addition,
as we discussed in this paper, these Killing vectors may be constructed through  solutions to the Schr\"{o}dinger-type equation $\Psi$ (see appendix \ref{Appen-A} for more
discussions). The $sl(2,R)$ doublet $\Psi$ is indeed what we would like to identify as the ``edge-state'' (this makes a very close parallel to the notion of edge state
in quantum Hall literature, e.g. see \cite{Edgestate-QHY}).  Explicitly, the action which describes the AdS$_3$ gravity and its edge-states is
\be\label{CS+Edge}
S=S_{\text{CS}}+S_{\text{edge}}\,,
\ee
where $S_{\text{CS}}$ is given in \eqref{CS-gravity} and
\be
S_{\text{edge}}=\frac{c}{6}\int dx^+dx^-\ \left(\bar\Psi D_+\Psi-\bar\Phi D_-\Phi\right)\,,
\ee
where $D_\pm=\partial_\pm+\mathcal{A}_\pm$ (\emph{cf.} Appendix \ref{sec-A-1}). The idea is then to quantize \eqref{CS+Edge}. Here, the edge states $\Psi$ and $\Phi$ play a fundamental role and boundary gravitons may be recovered through them.

Performing the details of this quantization is postponed to our upcoming work. However,
before closing we would like to discuss some other interesting and related ideas:
\paragraph{The extremal, degenerate horizon AdS$_3$ geometries and their quantization.} As we discussed in sections \ref{sec-3} and \ref{sec-4} there is an interesting
subclass of geometries of \eqref{generic-solutions} which correspond to cases with degenerate horizon. These geometries which are
obtained by setting $f_+=0$ were also constructed and discussed in \cite{LL-paper}. As discussed in the end of section \ref{sec-4}, these geometries
unlike the generic $f_+\neq 0$ case, have generically  event horizon, too. As pointed out in section \ref{sec-4}, one should modify our semiclassical microstate proposal a little bit
to fit these cases. Moreover, this case can come with compact or noncompact $x^+$, where the two are related through the DLCQ procedure in the dual CFT \cite{DLCQ-paper}.
Since in this case we are dealing with a single function $f_-$ and hence a chiral sector of a Virasoro algebra, it would be a very good test ground for examining the
quantization proposals outlined above.

\paragraph{Connection to Kerr/CFT.} The above picture for quantizing AdS$_3$ gravity will have direct implication and application in the higher dimensional gravity theories.
In particular, it is known that geometry appearing in Near Horizon limit of Extremal black hole Geometry (NHEG)
generically develop an AdS$_2$ throat with $Sl(2,R)\times U(1)^N$ isometry, for a review see \cite{KL-review}. The Kerr/CFT proposal \cite{Kerr/CFT},\footnote{The  laws of NHEG mechanics
and their derivation based on Neother-Wald conserved charged has been crystalized and stated in \cite{HSS}.} then states that
 microstates corresponding to the NHEG (and hence to the extremal black hole) is described by a chiral 2d CFT and that  states of this 2d CFT
are labeled by the asymptotic symmetry group associated with the diffeomorphisms of prescribed boundary conditions on the NHEG background.
It would be very interesting to  extend the picture for AdS$_3$ boundary gravitons we presented here to the NHEG boundary gravitons discussed in Kerr/CFT proposal
\cite{Kerr/CFT}.

\subsection*{Acknowledgement}

We would like to especially thank Joan Simon for his valuable insights and collaboration at some stages of this work. We would like to thank Glenn Barnich, Steve Carlip, James Lucietti for discussions and useful comments on the draft. We also thank Dongsu Bak and Soo-Jong Rey for useful discussions.
M.M.Sh-J. would like to thank the International Visiting Scholar program of of Kyung Hee University where part of this work was completed.

\appendix

\section{More on the algebra of Killings}\label{Appen-A}
A straightforward computation reveals that for any two  Killing vectors given by $K$ and $H$ functions,  $K'H-KH'$ is also forming a Killing;
more explicitly, Lie bracket of two Killings generated by $K$ and $H$ functions is a Killing generated by  $K'H-KH'$. This suggests the following
definition of  bracket among functions:
\be\label{bracket-function}
\{f,g\}_{{}_S}\equiv f'g-f g'\,,\qquad \forall\ f,g\,.
\ee
With the above definition and the normalization chosen for the functions $\psi_\alpha$,
\be
\{\psi_\alpha,\psi_\beta\}_{{}_S}=\epsilon_{\alpha\beta}\,.
\ee
The index $S$ on the bracket is chosen to denote the fact that this bracket is induced from the normalization condition of the Schr\"{o}dinger equation \eqref{normalization}.

Let us now compute the algebra of Killings $K_a, a=-,+,0$ generated by $\psi_1,\psi_2$, as given in \eqref{Ka}.
A simple calculation shows that
\be\label{Killing-algebra}
\begin{split}
\{K_{-1},K_{+1}\}_{{}_S}=2 K_0\,,&\qquad \{K_{0},K_{\pm 1}\}_{{}_S}=\pm K_{\pm 1}\,.
\end{split}
\ee
The above is clearly the $sl(2,R)$ algebra. In fact one can show that the above bracket between the Killings $K_a$ is nothing but the
Lie bracket of these vectors, see \cite{BT} for further discussions.

\subsection{Relation to 2-d Dirac equation}\label{sec-A-1}
From second order differential equations (\ref{Schrodinger}) on $\psi$ and $\phi$, one can get first rank equation similar to Dirac's equation.
Let us introduce doublets $\Psi$ and $\Phi$
\be
 \Psi=\left( \begin{array}{ccc}
\psi_1 \\
{\psi_2}
 \end{array} \right)\,, \quad\quad
  \Phi=\left( \begin{array}{ccc}
\phi_1 \\
{\phi_2}
 \end{array} \right),
 \ee
where $\psi_\alpha$ and $\phi_\alpha$ are solutions to \eqref{Schrodinger}. One can then show that $\Psi$ and $\Phi$ satisfy following equations
\be\label{Phi-Psi-Dirac}
{\partial}_+ \Psi +{\mathcal{A}}_+\Psi=0\,, \qquad    {\partial}_- \Phi + \mathcal{A}_-\Phi=0\,,
 \ee
with ``gauge fields'' $\mathcal{A}_\pm$
\be
  \mathcal{A}_\pm= A_\pm\mathds{1}+B_\pm n_a \tau^a\,,\qquad a=0,+,-\,,\qquad
\ee
where  $\tau^a$ are $2\times 2$ generators of $sl(2,R)$ given in \eqref{sl2R-2x2}, $n_a$ is such that $(n_a\tau^a)^2=\mathds{1}$ and
\be
\frac{\partial B}{B}=2A\,,\qquad -\partial A+A^2+ B^2=f\,,
\ee
(with similar relations for both $+$ and $-$ cases).  As we see the gauge field is specified by a single function $B$ (the $sl(2,R)$ gauge field)
which in turn is fixed by the function in the metric $f$. The $sl(2,R)$ part of the gauge field is along a unit vector $n_a$, which specifies a
point on a unit radius AdS$_2$ space \cite{HSS}.


With the above one can write the explicit solution for \eqref{Phi-Psi-Dirac} as
\be
\Psi=P\exp(-\int_{x^+} \mathcal{A}_+)\ \Psi_0\,,\qquad \Phi=P\exp(-\int_{x^-} \mathcal{A}_-)\ \Phi_0\,,
\ee
where $\Psi_0$ and $\Phi_0$ are two arbitrary constant $Sl(2,R)$ doublets and $P$ denotes the path ordering.

\subsection{Representation of $sl(2,R)$ in terms of $\Psi$-bilinears}

One can give a representation of the algebra of Killings \eqref{Killing-algebra} using the $\Psi$-doublets.
To this end we choose the $(0,-,+)$ basis for $sl(2,R)$ and note that in $2\times 2$ representation of $sl(2,R)$ algebra, generators take the form
\be\label{sl2R-2x2}
\tau^{0}=\frac12\left( \begin{array}{cc}
1 & 0\\
0 & -1
 \end{array} \right)\,,\qquad
\tau^{+}=\left( \begin{array}{cc}
0 & 1\\
0 & 0
 \end{array} \right)\,,\qquad
\tau^{-}=\left( \begin{array}{cc}
0 & 0\\
-1 &0
 \end{array} \right)\,.
\ee
The $sl(2,R)$ metric $g^{ab}$ is then $\{\tau^a,\tau^b\}=2g^{ab}\mathds{1}$. One can represent the Killing generators $K^a$ given in \eqref{Ka} in a more inspiring and formal way as
\be
K^a=\bar\Psi \tau^a\Psi\,,
\ee
where
\be
 \Psi=\left( \begin{array}{ccc}
\psi_1 \\
{\psi_2}
 \end{array} \right)\,,
 \ee
is an $sl(2,R)$ doublet and
\be
\bar\Psi=\Psi^T\gamma\,,\qquad \gamma=\frac12\left( \begin{array}{cc}
0 & 1\\
-1 & 0
 \end{array} \right)\,.
\ee


\section{A brief review on Hill's Equation and Floquet theorem}\label{Appen-B}

\emph{Note: Unless another reference is given, all results in this appendix are taken from \cite{Hill-Eq-Book}.}

Equation \eqref{Schrodinger} with periodic $f$ is called Hill's equation. For definiteness, consider
\be\label{Hill-Equation}
\psi''+(\lambda+Q(x))\psi=0\,,
\ee
where
\be
Q(x+2\pi)=Q(x)\,,\qquad \int_0^{2\pi} dx\  Q(x)=0\,.
\ee
Let us denote the two basic solutions to the Hill's equation by $\psi_1$ and $\psi_2$. Given the linearity of Hill's equation \eqref{Hill-Equation}
one can always choose the initial and normalization conditions such that
\be\label{initial-condition}
\psi_1(0)=1\,,\qquad \psi_2(0)=0\,,\qquad \psi_1'(0)=0\,,\qquad \psi_2'(0)=1\,.
\ee

Although Hill's equation for a given periodic $Q(x)$ has periodic solutions for specific (discrete set) of $\lambda$'s, periodicity of $Q(x)$ does not imply
periodicity of $\psi_1,\ \psi_2$. Nonetheless one can show that
\be\begin{split}
\psi_1(x+2\pi)&= \psi_1(2\pi) \psi_1(x)+\psi_1'(2\pi) \psi_2(x)\,,\cr
\psi_2(x+2\pi)&= \psi_2(2\pi) \psi_1(x)+\psi_2'(2\pi) \psi_2(x)\,,
\end{split}\ee
which implies    ``quasi-periodicity'' condition on $\psi$'s, stated in Floquet theorem:

\emph{Solutions of Hill's equation satisfy the following Floquet behavior,}
\be\label{Floquet}
\psi_1(x)= e^{i\nu x} P_\nu(x)\,,\ \psi_2(x)= e^{-i\nu x}  P_2(x)\,, \qquad P_1(x+2\pi)=P_1(x),\ P_2(x+2\pi)= P_2(x)
\ee
\emph{and $e^{2i\pi\nu}$, $-1/2< Re(\nu)\leq1/2$  is the eigenvalue of Wronskian matrix}
\be
\left(\begin{array}{cc} \psi_1(2\pi) & \psi_2(2\pi) \cr \psi_1'(2\pi) & \psi_2'(2\pi)
\end{array}\right),
\ee
\emph{$\nu$ is called the Floquet exponent and one can show that $2\cos(2\pi\nu)=\psi_1(2\pi)+\psi_2'(2\pi)$.}

Note that in general the Floquet exponent $\nu$ is a function of $\lambda$. Moreover, one can show that
$$
\psi(x+2\pi)+\psi(x-2\pi)=2\cos(2\pi\nu)\psi(x)\,.
$$

A useful indicator of solutions of Hill's equation is the \emph{discriminant} $\Delta(\lambda)$,
\be
\Delta(\lambda)=\psi_1(2\pi)+\psi_2'(2\pi)\,.
\ee
It has been shown that $\Delta$ is an entire function of order 1/2 of $\lambda$.
For a given Floquet exponent $\nu$, $\Delta(\lambda)-2\cos(2\pi\nu)=0$ has infinitely many roots $\lambda$. Conversely, for a given $\lambda$, one may compute $\nu$
once we have $\Delta(\lambda)$. The discriminant $\Delta$ can be expressed as an infinite determinant involving the Fourier coefficients of $Q(x)$
\cite{Hill-Eq-Book}.

\subsection{Periodic solutions, spectrum of eigenvalues and (in)stability intervals}

Although solutions of Hill's equation for generic ``eigenvalues'' $\lambda$ are not periodic, there are specific values of $\lambda$ for a given $Q(x)$
for which we have periodic solutions. It is readily seen that periodic solutions are those with $\nu=0$ or $\nu=1/2$. It has been shown that
\bn
\item Eigenfunctions which are periodic with period $2\pi$ correspond to $\nu=0$ case and form a discrete set.
Let us denote them by $\psi_k$, $k=0,1,\cdots$, i.e. $\psi_k(x)=\psi_k(x+2\pi)$ and the corresponding eigenvalues by $\lambda_k$.
We choose $k=0$ for the ``ground state'' with smallest value of $\lambda$.\vskip 1mm
\item The explicit values of eigenvalues $\lambda_k$ is completely specified, and can in principle be computed, given
the potential $Q(x)$, through the ``Hill's determinant.''\vskip 1mm
\item For generic ``potential'' $Q(x)$, we  have an infinite number of periodic solutions. However, there are specific $Q(x)$ where there are finite number of
periodic solutions \cite{McKaen}.\vskip 1mm
\item The ground state $\psi_0$ has no zeros and $\chi=\frac{\psi_0'}{\psi_0}$ is a smooth function averaging to zero (in the $[0,2\pi]$ range).
One can then show that \cite{ground-state-bound}
\be
\lambda_0=-\frac{1}{2\pi}\int_0^{2\pi} \chi^2 \geq -\frac{1}{2\pi}\int_0^{2\pi} Q^2\,.
\ee
That is, the ground state energy $\lambda_0$ is always negative and has a lower bound.\vskip 1mm
\item This equation for $\nu=1/2$ case has also solutions which are periodic with period $4\pi$.
Let us denote these by $\tilde\psi_k$, where $\tilde\psi(x)=\tilde\psi(x+4\pi)$ and
the  corresponding eigenvalues  by $\mu_k$, with $k=1,2,\cdots$. \vskip 1mm
\item It has been proved that for every $2\pi$ periodic solution there is a $4\pi$ periodic solution, except for the ``ground state'' and that the eigenvalues are ordered in the following way:
\be
\lambda_0< \mu_1\leq \mu_2 <\lambda_1 \leq \lambda_2< \mu_3\leq \mu_4 <\lambda_3 \leq \lambda_4< \ \cdots
\ee
\vskip 1mm
\item Stable solution (those with bounded eigenfunctions) are those with \emph{real} $\nu$ which happens if
$$
|\Delta(\lambda)|< 2\,.
$$
(Note that $\Delta$ is real, $\rho=e^{2i\pi\nu}$ and $\rho^2-\rho\Delta+1=0$.) The stability condition is satisfied
if $\lambda$ is in the stability bands:
$$
\lambda\in (\lambda_0, \mu_1)\cup (\mu_2,\lambda_1)\cup
(\lambda_2, \mu_3)\cup \cdots.$$
The compliment of the stability intervals are the instability bands (where the eigenfunction grow exponentially), $(-\infty,\lambda_0)\cup
(\mu_1,\mu_2)\cup(\lambda_1,\lambda_2)\cdots $. In the instability intervals one of the solutions has exponential growth while the other is exponentially decaying.

Note that  $\lambda=\lambda_k$ or $\mu_k$, respectively  corresponds to $\Delta=2$ ($\nu=0$) or $\Delta=-2$ ($\nu=1/2$). In this case the solutions are ``meta-stable;'' they do not fall-off exponentially.
These cases would be of interest to us.
\vskip 1mm
\item If the potential $Q(x)$ is infinitely differentiable and smooth then,
\be\label{lambda-mu-large-k}
\begin{split}
\lambda_{2k}&\simeq \lambda_{2k-1}=4k^2+\frac{1}{32\pi k^2} \int_0^{2\pi}{Q^2}+{\cal O}(\frac{1}{k^2})\,,\cr
\mu_{2k}&\simeq \mu_{2k-1}=(2k-1)^2+\frac{1}{32\pi k^2} \int_0^{2\pi}{Q^2}+{\cal O}(\frac{1}{k^2})\,.
\end{split}
\ee
Moreover, if $Q$ is smooth and infinitely differentiable, $\lambda_{2k}-\lambda_{2k-1}\sim e^{-k}$ and also  $\lambda_{2k}>(2k)^2$ for large $k$.
\vskip 1mm

\item The eigenvalues have the following recursion relations:
\be
\lambda_0+\sum_{k=1}^\infty (\lambda_{2k-1}+\lambda_{2k}- 8k^2)=0\,,\qquad \sum_{k=1}^\infty (\mu_{2k-1}+\mu_{2k}- 2(2k-1)^2)=0\,.
\ee
\en

\section{More on the degenerate horizon case}\label{LLvsUS}

To see how metric \eqref{f+=0-metric} and the metric in eq.(20) of \cite{LL-paper}:
\be\label{LL-metric}
ds^2=2dv\left[d\lambda +\frac{2}{\ell}\lambda(1+\frac12\lambda\gamma(x))dx\right]+(1+\lambda\gamma(x))^2dx^2,
\ee
are related, consider the following coordinate transformation
\be
\lambda=r-\frac{1}{\gamma(x)}\,.
\ee
Then a simple algebra reveals that \eqref{LL-metric} takes the form
\be\label{LL-metric-2}
ds^2=F(r,x)\left[dv+ \frac{dr}{F(r,x)}\right]^2+\frac{dr^2}{F(r,x)}+ r^2\left[(\gamma dx+\frac{dv}{\ell})-\frac{\beta(x)}{r^2}\frac{dv}{\ell}\right]^2
\ee
where
\be\label{F-beta}
F(r,x)=\frac{(r^2-\beta(x))^2}{\ell^2 r^2}\,,\qquad \beta(x)=\frac{1}{\gamma^2}-\frac{\ell \gamma'}{\gamma^3}\,,
\ee
where \emph{prime} denotes derivative w.r.t. $x$. It is readily seen that for constant $\gamma$ (which means constant $\beta$), the above metric \eqref{LL-metric-2} is the metric for an extremal BTZ with horizon radius
$r_h^2=\beta$. Note also that we should choose periodicity of $x$ coordinate, which is taken to be $R$ in \cite{LL-paper}, such that $\int_0^{2\pi R} \gamma dx=2\pi$.

\paragraph{Connection to our Schr\"{o}dinger eq.} From the above form one learns that a coordinate transformation like
\be\label{x-gamma-x+}
x^-=\int \gamma dx +K(\beta, r)\,,\qquad x^+=\frac{v}{\ell}+H(\beta,r)\,,
\ee
relates \eqref{LL-metric} to \eqref{f+=0-metric}, where the large $r$ expansion of $K$ and $H$ functions is given in eq.(30) of \cite{LL-paper}. The important and interesting
point is then how the function $\gamma(x)$ and our Schr\"{o}dinger equation wavefunctions are related. Here is the answer:
\be
\frac{d^2}{dx^-{}^2}\phi-f_-(x^-)\phi=0\quad \Longrightarrow \quad \frac{d}{dx^-}U+U^2=f_-\,,\qquad \phi\equiv \exp{(\int U)},\quad U=\frac{d\phi/dx^-}{\phi}\,.
\ee
Then \eqref{x-gamma-x+} implies that the derivative w.r.t. $x^-$ and derivative w.r.t. $x$ have a factor of $\gamma$ difference.
Noting this fact we can rewrite $\beta$ (\emph{cf.}\eqref{F-beta}) as
\be
\beta(x^-)=\frac{1}{\gamma^2}-\frac{1}{\gamma^2}\frac{d\gamma}{d x^-}.
\ee
On the other hand, comparison of asymptotic expansions of \eqref{LL-metric} and \eqref{f+=0-metric} reveals that
\be
\beta(x^-)=f_-(x^-)\,,
\ee
and the above in turn implies that
\be
U=\frac{1}{\gamma}\,.
\ee
In other words, $\gamma$ is directly related to the $log$-derivative of our wavefunction. It is intriguing that the same expression appears in our analysis for norm of
the Killing vectors (and location of horizons).

\paragraph{The Killing vectors.} Metrics \eqref{LL-metric} and \eqref{f+=0-metric} are indeed the same geometry written in two different coordinate systems. Since the coordinate
transformations which map the two are complicated and hence not illuminating, we find it more useful to compare the Killing vectors fields in these two coordinate systems.
As all  locally AdS$_3$ geometries, metrics \eqref{LL-metric} or equivalently \eqref{f+=0-metric} have local $sl(2,R)\times sl(2,R)$ isometries. To see this explicitly
we write the six Killing vector fields in both coordinate systems.
\bi
\item$\boldsymbol{sl(2,R)_v.}$ Let us start with the $sl(2,R)$ of \eqref{LL-metric} which involves the explicit Killing vector $\partial_v$:
\be\label{sl(2,R)v}
\begin{split}
K^v_-&=\frac{\ell}{\sqrt2}\partial_v \,,\cr
K^v_0&= \mathbf{X^+}\partial_v-\frac{1}{2}\frac{1}{1+\lambda\gamma}\bigl(\lambda\partial_\lambda-\ell\partial_x\bigr)-\frac12\partial_\lambda \,,\cr
K^v_+&= {\sqrt2}\left[\frac{(\mathbf{X^+})^2}{\ell}\partial_v-\frac{\mathbf{X^+}}{1+\lambda\gamma}\bigl(\lambda\partial_\lambda-\ell\partial_x\bigr)-
\ell(\frac12+\frac{\lambda}{\ell}\frac{\mathbf{X^+}}{\ell})\partial_\lambda
 \right] \,,
\end{split}
\ee
where
\be\label{Gamma-X+}
\mathbf{X^+}=v+\frac{\ell}{2}\Gamma,\quad \frac{d\Gamma}{dx}\equiv\gamma(x)\,.
\ee
This is to be compared with $sl(2,R)_+$ of \eqref{f+=0-metric} which involves $\partial_{x^+}$ as one of generators, i.e.
\be\label{sl(2,R)+}
\begin{split}
K^+_-&=\partial_{x^+} \,,\cr
K^+_0&=x^+\partial_{x^+}-\frac{r}{2}\partial_r  \,,\cr
K^+_+&=(x^+)^2\partial_{x^+}-{x^+}{r}\partial_r+\frac{\ell^4}{r^4} f_-(x^-)\partial_{x^+}+\frac{\ell^2}{r^2}\partial_{x^-}  \,.
\end{split}
\ee
Let us now analyze periodicity of the \eqref{sl(2,R)v} generators. In order these Killing vectors to be well-defined we should require $1+\lambda\gamma$ be non-vanishing everywhere for $\lambda\geq 0$. This in turn implies that $\gamma\geq 0$. Except for the $\gamma=0$ case which corresponds to SDO, the function $\Gamma$ and hence $\mathbf{X^+}$ defined in \eqref{Gamma-X+}, will not be periodic in $x$. Therefore, in general only $K^v_-$ (among $sl(2,R)_v$ generators) is periodic. (Note also that
only $K^v_-$ is  also periodic in $v$.) Therefore, as pointed out in
the discussions of section \ref{sec-4},  only $U(1)_v$ part of $sl(2,R)_v$ is the global isometry of geometries in \cite{LL-paper}. This is in accord with the analysis in that paper. As the notation suggests, the coordinates $v$ and $x^+$ are related through a coordinate transformation of the form $\ell x^+=\mathbf{X^+}+H(\lambda, x)$.\\

\item$\boldsymbol{sl(2,R)_x.}$ The other $sl(2,R)$ local isometry of \eqref{LL-metric}, denoted by $sl(2,R)_x$ is given by
\be\label{sl(2,R)x}
\begin{split}
K^x_-&=\frac{\sqrt2}{2}e^{\frac{2x}{\ell}}\left[\frac{\ell^2}{2}\partial_v+\frac{\lambda}{1+\lambda\gamma}(\lambda\partial_\lambda-{\ell}\partial_x)\right] \,,\cr
K^x_0&= -\frac12\left[\ell\mathbf{Y}\partial_v+\frac{2\frac{\lambda}{\ell}\mathbf{Y}-1}{1+\lambda\gamma}(\lambda\partial_\lambda-{\ell}\partial_x)+\lambda\partial_\lambda\right] \,,\cr
K^x_+&= {\sqrt2}e^{\frac{-2x}{\ell}}\left[\frac12\mathbf{Y}^2\partial_v+\frac{\mathbf{Y}(\frac{\lambda}{\ell}\mathbf{Y}-1)}{\ell(1+\lambda\gamma)}
(\lambda\partial_\lambda-{\ell}\partial_x)+(\frac{\lambda}{\ell}\mathbf{Y}-1)\partial_\lambda\right] \,,
\end{split}
\ee
where $\mathbf{Y}$ is the \emph{periodic} solution to
\be
\mathbf{Y}'-\frac{2}{\ell}\mathbf{Y}=\gamma\,.
\ee
(This may be compared with eq.(15) of \cite{LL-paper}.)  $sl(2,R)_x$ is the counterpart of $sl(2,R)_-$ of \eqref{f+=0-metric} given in \eqref{hi-killing}, i.e.
\be\label{sl(2,R)-}
K^a_-=K_a\partial_{x^-} +\frac{\ell^2}{2r^2}K_a''\partial_{x^+}-\frac{r}{2}K'_a\partial_r \,,
\ee
where $K_a$ are as given in \eqref{Ka}, explicitly
$$
K_{-1}=\frac12\phi_1^2\,,\quad K_0=\frac12\phi_1\phi_2\,,\quad K_{+1}=\frac12\phi_2^2\,,
$$
and $\phi_\alpha$ are the two normalized linearly independent solutions to $\phi''-f_-\phi=0$ with $\phi_1'\phi_2-\phi_1\phi'_2=1$.
\ei
As one can explicitly see from \eqref{sl(2,R)x}, $K^x_0$ is periodic in $x$ while $K^x_\pm$ are not periodic. This is in perfect agreement with our earlier discussions on
the solutions for the Hill's equation and the Floquet theorem, which states that $K^0_-$ in \eqref{sl(2,R)-} is periodic, while $K^\pm_-$ are exponentially decaying or growing with $x$.

To summarize, the metric \eqref{LL-metric} (with noncompact or compact $v$) with $\gamma\neq 0$ has global $U(1)_v\times U(1)_x$ isometry as the metric in \eqref{f+=0-metric} (with noncompact or compact $x^+$).
In the noncompact $x^+$ (or $v$) case these geometries correspond to Brown-Henneaux boundary gravitons of AdS$_3$ self-dual orbifold, while when $x^+$ (or equivalently $v$) is compact,
the geometry corresponds to extremal BTZ geometry excited with Brown-Henneaux boundary gravitons.


{}

\end{document}